\documentclass{article}
\newcommand{\ds }{\displaystyle}

\newcommand{\be}{\begin{equation}}
\newcommand{\ee}{\end{equation}}
\newcommand{\bea}{\begin{eqnarray}}
\newcommand{\eea}{\end{eqnarray}}
\newcommand{\ci}{\cite}
\newcommand{\bi}{\bibitem}
\newcommand{\nono}{\nonumber \\}

\newcommand{\dd}{\partial}

\newcommand{{\bfna}}{\mbox{\boldmath$\vec{\nabla}$}}

\begin{document}

\title{\sl A theorem for the normalization of continuous spectrum stationary states}
\vspace{1 true cm}
\author{G. K\"albermann$^*$
\\Soil and Water dept.(Emeritus), Faculty of
Agriculture, Rehovot 76100, Israel}

\maketitle

\vspace{3 true cm}
\begin{abstract}
We present analytic formulae that simplify the evaluation of the normalization of continuous spectrum stationary states
in the one-dimensional Schr\"odinger equation. 
\end{abstract}
\vspace{3 true cm}


Subjects: orthonormality, scattering states

\vspace{3 true cm}
$^*${\sl e-mail address: germankalb@gmail.com}

\newpage
\section{\sl Introduction}

The normalization of stationary states continuous spectrum one dimensional Schr\"odinger wave functions
in the presence of a real and finite range potential often becomes involved and cumbersome.
This normalization is essential for the use of these waves in a perturbative solution
of more complicated problems in which extra interactions are present. It is also essential
for the expression of completeness of the spectrum to have properly normalized wave functions.\ci{amaku},\ci{
dalab} We here show that the calculation gets simplified resorting to a hitherto unknown mathematical equalities.

In section 2 we derive these relations. Section 3 applies one of the equalities to the case of a square well. Section 4
evaluates the normalization of continuous spectrum wave functions.

\section{\sl Derivation of the equalities}

The Schr\"odinger equation for the one dimensional system labeled by a mass $\ds m$
with arbitrary potential V(x) is 

\be\label{sch1}
i\frac{\dd\Psi}{\dd t}=\frac{-1}{2~m}\frac{\dd^2 \Psi}{\dd x^2}+
V(x)\Psi
\ee

For stationary states
\bea\label{sch2}
i\frac{\dd\Psi}{\dd t}=E\psi=\frac{-1}{2~m}\frac{\dd^2 \psi}{\dd x^2}+
V(x)\psi
\eea

with E, the eigenvalue of the energy of the state, positive for the continuous case 
and negative for discrete bound states.

Using eq.(\ref{sch2}) we can readily see that stationary state waves can be taken as real
except for the trivial time evolution factor $\ds \Psi(x,t)=e^{-iEt}\psi(x)$.
The full spectrum is spanned by even and odd spatial symmetry functions.
For the continuous spectrum  labeled by a wavenumber $\ds k$ with $\ds E=\frac{k^2}{2m}$,
the spectrum is exhausted by taking even and odd waves with $\ds k \ge 0$

Consider the following integral

\bea\label{int1}
I=\int_{x_1}^{x_2}\psi_k(x)~\psi_{k'}(x) dx
\eea

\noindent Using eq.(\ref{sch2}) we obtain

\bea\label{int2}
I=\int_{x_1}^{x_2}\left( \frac{-1}{k^2}\frac{\dd^2 \psi}{\dd x^2}+
\frac{V(x)}{E}\psi_k(x)\right)\psi_{k'}(x) 
\eea

Integration by parts and repeated use of eq.(\ref{sch2}) yield

\bea\label{int3}
I=\frac{1}{\left(k^2-k'^2\right)}\left(\frac{\dd\psi_{k'}}{\dd x}~\psi_k(x)
-\frac{\dd\psi_k}{\dd x}~\psi_{k'}(x) \right)\bigg|_{x_1}^{x_2}
\eea

We will apply eq.(\ref{int3}) to the normalization
of continuous spectrum wave functions. The expression is well defined for $\ds k\ne k'$. 
The case of $\ds k=k'$ will yield a $\delta$ function normalization naturally.

Consider eq.(\ref {int1}) along the whole real line, the normalization integral

\bea\label{int4}
\mathit {I} &=\mathit{I}_1+\mathit{I}_2+\mathit{I}_3\nono
\mathit{I} &=\int_{-\infty}^{\infty}\psi_k(x)~\psi_{k'}(x) dx\nono
\mathit{I}_1 &=\int_{-\infty}^{x_a}\psi_k(x)~\psi_{k'}(x) dx\nono
\mathit{I}_2 &=\int_{x_a}^{x_b}\psi_k(x)~\psi_{k'}(x) dx\nono
 \mathit{I}_3 &= \int_{x_b}^{\infty}\psi_k(x)~\psi_{k'}(x) dx
\eea

\noindent where $\ds x_a $, $\ds x_b $ are the left and right boundaries of the finite range potential V(x),
or alternatively any chosen points outside the potential range at which the potential is negligible.
For the the first and third pieces of the integral of eq.(\ref{int4}) we can take the asymptotic solutions for the 
potential free region. For $x\ge0$

\bea\label{asymptotic}
\psi^{out}_k(x)=A(k)~e^{ikx}+A^*(k)e^{-ikx}
\eea

\noindent and similar expressions for $\ds x\le 0$ appropriate for the even-odd character of the wave.

For the interior region we have need

\bea\label{int5}
\mathit{I}_2=\frac{1}{\left(k^2-k'^2\right)}\left( 
\frac{\dd\psi^{int}_{k'}}{\dd x}~\psi^{int}_k(x)~-\frac{\dd{\psi^{int}
_k}}{\dd x}~\psi^{int}_{k'}(x) \right)\bigg|_{x_a}^{x_b}
\eea

where $\psi^{in}$ is the solution of the Schr\"odinger equation for the region where the potential
is nonvanishing, referred to as the interior region.
In equation \ref{int5} both the numerator and the denominator vanish when $\ds k=k'$. As shown
below, the final expression is well defined, leading to the standard normalization of continuum states in terms of the Dirac $\ds \delta$
function.

Continuity of the wave function and its derivative at $\ds x_a $ and $\ds x_b$
demands

\bea\label{boundary}
\psi^{int}_k(x_a)=&~\psi^{out}_k(x_a)\nono
\psi^{int}_k(x_b)=&~\psi^{out}_k(x_b)\nono
\frac{\dd\psi^{int}_k}{\dd x}(x_a)&=~\frac{\dd\psi^{out}_k}{\dd x}(x_a)\nono
\frac{\dd\psi^{int}_k}{\dd x}(x_b)&=~ \frac{\dd\psi^{out}_k}{\dd x}(x_b)
\eea

where $\psi^{out}$ is the solution of the Schr\"odinger equation for the region where the potential effectively
vanishes, the outer region. This comes as no surprise do the property of the Wronskian appearing in eq.(\ref{int5}).

Therefore, we can replace $\ds \psi^{int}$ by $\ds \psi^{out}$ in eq.(\ref{int5}).
The normalization proceeds solely through the knowledge of $\psi^{out}$.

The outer function $\ds \psi^{out}$ obeys the the Schr\"odinger eqation without a potential, namely

\bea\label{sch3}
i\frac{\dd\Psi}{\dd t}=E\psi=\frac{-1}{2~m}\frac{\dd^2 \psi}{\dd x^2}
\eea

It does not solve the equation in the inner region bearing a potential. However, it obeys the boundary
conditions of eq.(\ref{boundary}). Both properties imply, that although it is a continuous function even
extended to the inner region, it's derivative is not. There are points $\ds x_i$ , at least one, at which the derivative is discontinuous.
In the next section we exemplify this characteristic for the case of a square well.

The solution of the outer region applied to the inner region reads

\bea\label{outsol}
\psi^{out}(x)=\sum_i\bigg(\Theta(x_i-x) \psi_1(x)+\Theta(x-x_i) \psi_2(x)\bigg)
\eea

With $\Theta$, the step function. 

Continuity of the function at $\ds x_i$ implies

\bea\label{outsolder}
\psi'~^{out}(x)=\sum_i\bigg(\Theta(x_i-x) \psi'_1(x)+\Theta(x-x_i) \psi'_2(x)\bigg)
\eea

where $\ds \delta$ is the Dirac $\ds \delta$ function and primes denote derivatives with respect to x.

The second derivative reads

\bea\label{outsoldd}
\psi''~^{out}(x) =\sum_i\bigg(\Theta(x_i-x) \psi''_1(x)+\Theta(x-x_i) \psi''_2(x)+\delta(x-x_i)\big(\psi'_2(x)- \psi'_1(x)\big)\bigg)
\eea

Applying the boundary conditions of eq.(\ref{boundary}) to eq.(\ref{int5}) we obtain

\bea\label{int51}
\mathit{I}_2=\frac{1}{(k^2-k'^2)}\left( 
\frac{\dd\psi^{out}_{k'}}{\dd x}~\psi^{out}_k(x)~-\frac{\dd{\psi^{out}
_k}}{\dd x}~\psi^{out}_{k'}(x) \right)\bigg|_{x_a}^{x_b}
\eea

using eq.(\ref{sch3}), eq.(\ref{int51}) becomes

\bea\label{int52}
\mathit{I}_2=\frac{1}{\left(k^2-k'^2\right)}\int_{x_1}^{x_2}\left( 
\frac{-\dd^2\psi^{out}_{k'}}{\dd x^2}~\psi^{out}_k(x)~+\frac{\dd^2{\psi^{out}
_k}}{\dd x^2}~\psi^{out}_{k'}(x) \right)
\eea
Inserting eq.(\ref{outsoldd} ) into eq.(\ref{int52}) yields

\bea\label{int6}
&\mathit{I}_2=\int_{x_1}^{x_2}\psi^{int}_k(x)~\psi^{int}_{k'}(x) dx\nono
&=\int_{x_1}^{x_2}\psi^{out}_k(x)~\psi^{out}_{k'}(x) dx
+\sum_i\bigg(\big(\psi'_{2,k'}(x_i)-\psi'_{1,k'}(x_i)\big)\psi_k(x_i)-\big(\psi'_{2,k}(x_i)-\psi'_{1,k}(x_i)\big)\psi_k'(x_i)\bigg)
\eea

Eq.(\ref{int6}) is a new identity that connects inner and outer regions normalization integrals. 
In the next section we show the validity of eq.(\ref{int6}) for the special case of a  square well.

\section{\sl The case of a square well}

We here evaluate eq.(\ref{int6}) explicitly.
Consider a particle of mass $\ds m$ in a region of space having a square well potential of strength $\ds -V_0$, placed between $\ds x=-d$ and $\ds x=d$. 
A stationary continuous spectrum even wave function for $\ds -d\le x\le d$ reads 

\bea\label{evenin}
\Psi^{in}=  cos(q x) 
\eea

whereas for $\ds |x| \ge d$

\bea\label{evenout}
\psi^{out}= a cos(k|x|+\phi)
\eea

The wave function derivative of eq.(\ref{evenout}) is discontinuous at the origin.

Where $\ds \frac{k^2}{2 m} = \frac{q^2}{2 m}-V_0$.
The overall normalization factor is irrelevant for the calculation.

The boundary conditions for the function and the derivative at $\ds |x|=d$ are

\bea\label{boundary1}
cos(kd)=& a cos(kd+\phi)\nono
q sin(qd)=& a k sin(kd+\phi)
\eea

Inserting eq(\ref{evenin}) into eq.(\ref{int6}) we find 

\bea\label{int61}
&\mathit{I}_2=\int_{-d}^{d}\psi^{int}_k(x)~\psi^{int}_{k'}(x) dx\nono
&=\frac{1}{q^2-q'^2}\bigg(q sin(qd) cos(q'd)-q' sin(q'd) cos(q d)\bigg)
\eea

where $\ds q'$ corresponds to $\ds k'$.

For the integral of eq.(\ref{int6}), in terms of the the wave function of eq.(\ref{evenout})
we use

\bea\label{evenout1}
\psi^{out}= a cos(kx+\phi)\Theta(x)+ a cos(-kx+\phi)\Theta(-x)
\eea

The derivative of the function is discontinuous at $\ds x=0$.
The expression of eq.(\ref{int6}) in terms of $\ds \psi^{out}$ including the contributions at the discontinuity at $x=0$ becomes

\bea\label{int62}
\mathit{I}_2=\frac{1}{k^2-k'^2}\bigg(k sin(kd+\phi) cos(k'd+\phi')-k' sin(k'd+\phi') cos(k d+\phi)\bigg)
\eea

Inserting eq.(\ref{boundary1}) and $\ds k^2-k'^2=q^2-q'^2$ into eq.(\ref{int62}) we find that eq.(\ref{int62}) is identical to
eq.(\ref{int61}). Hence, eq.(\ref{int6}) is correct.

The case of odd solutions is straightforward also.

\section{\sl Normalization of continuous spectrum wave functions}

The normalization integral of eq.(\ref{int4}) can be now evaluated resorting to eq.(\ref{int3}) or eq.(\ref{int6})
 
In both cases only $\ds\psi^{out}$ is needed for the calculation. 

Using eq.(\ref{asymptotic}) in eq.(\ref{int3}) we  obtain 

\bea\label{result}
\mathit{I}= 4\pi~|A|^2\delta(k-k')
\eea

Knowing the continuous spectrum wave function and we can normalize the function to yield the
properly normalized continuous spectrum stationary state wave functions $\ds\psi^{norm}$
to be

\bea\label{normalized}
\psi^{norm} =\frac{\psi}{2\sqrt{\pi}|A|}
\eea

\end{document}